\documentclass{doublecol-new}

\usepackage{graphicx}
\usepackage[numbers]{natbib}
\usepackage{mathrsfs}

\theoremstyle{TH}{

}

\theoremstyle{THrm}{

}

\theoremstyle{THhit}{

}

\makeatletter
\def\theequation{\arabic{equation}}

\makeatother

\begin{document}%

\setcounter{page}{1}

\LRH{F. Zhang}

\RRH{Improving the quantification of overshooting shock-capturing oscillations}
\VOL{x}

\ISSUE{x}

\PUBYEAR{xxxx}

\BottomCatch


\PUBYEAR{2023}

\subtitle{}

\title{Improving the quantification of overshooting shock-capturing oscillations}

\authorA{Fan Zhang} 
\affA{Centre for mathematical Plasma-Astrophysics, Department of Mathematics, KU Leuven, Leuven 3001, Belgium \\
Institute of Theoretical Astrophysics, University of Oslo, PO Box 1029 Blindern, 0315 Oslo, Norway\\
Rosseland Centre for Solar Physics, University of Oslo, PO Box 1029 Blindern, 0315 Oslo, Norway\\
E-mail: fan.zhang@astro.uio.no}

\begin{abstract}
An approach for quantitatively evaluating overshooting oscillations is designed to characterize the performance of shock-capturing schemes. Specifically, following our previous work focused on cases with only discontinuities, now we account for the concurrent presences of discontinuities and smooth waves, each with a complete set of supported modes on a given discretization. The linear advection equation is taken as the model equation, and a standardized measurement is given for overshooting oscillations produced by shock-capturing schemes. Thereby, we can quantitatively reveal the shock-capturing robustness of, for example, weighted essentially non-oscillatory schemes, by comparing and analyzing the resulting overshoots. In particular, we are able to find out the ranges of wavenumbers in which the numerical schemes are especially prone to produce overshooting oscillations. While lower dissipation is usually anticipated for high-order schemes, we provide a simple measurement for evaluating the shock-capturing robustness, which was not trivial due to the nonlinearity of shock-capturing computations.
\end{abstract}

\KEYWORD{Shock-capturing; overshooting oscillation; quantitative measurement; high-order scheme; finite-difference method.}


\maketitle
\section{Introduction}
Realistic compressible flows studied in (aero)space and astrophysical research frequently involve interactions between different types of flow features such as vortices, shock waves, etc. \cite{Drikakis_2003,Wedemeyer_2004,Knight_2012,Jiang_2015}.   Therefore, the capability of resolving multi-scale smooth structures  and meanwhile capturing shock waves is usually desired for modeling complex flows. However, while trying to  achieve high-order numerical accuracy and low numerical dissipation for shock-capturing schemes, it is  challenging to retain robustness at the same time. Frequently, either the high-order accurate methods would be used for less challenging scenarios \cite{Zhang_Zhang_2019,Xu2021}, or numerical oscillations  need to be removed by relatively diffusive  mechanisms  \cite{Balsara2000,Hu2013} that could deteriorate the accuracy and resolution in certain scenarios \cite{Zhao2019}. 

The difficulty of balancing robustness and accuracy results from, at least partly, lacking comprehensive quantitative evaluation approaches for overshooting shock-capturing oscillations that may have  computations terminated by causing negative values in scalar variables, or may potentially contaminate numerical results \cite{zhangcicp_2018,Xu2021}. 
Without a quantitative evaluation for examining the robustness of a shock-capturing scheme, typically one may perform a variety of numerical simulations involving discontinuities \cite{Lax1998}, and the failure or success of a simulation is considered as an indicator of shock-capturing robustness.
However, the failure or success of a simulation does not answer in what specific situation a numerical scheme may fail, or may not. Moreover, it is difficult, if not impossible, to compare the overshooting errors caused by different numerical schemes in complex simulations, where exact solutions are certainly absent.   {As a result, while   accuracy and resolution can be  evaluated qualitatively and even quantitatively, it is difficult to  predict or judge how (much)  robustness may be improved or deteriorated after having a different process implemented in simulations.}

Therefore, starting from the next section, we introduce a novel quantitative evaluation  process for the shock-capturing overshooting oscillations. We further  characterize the interactions between \textit{multiple} discontinuities and also \textit{multiple} smooth waves that have different amplitudes, while previously only a single discontinuity interacting with smooth waves \cite{Zhao2019} or  multiple discontinuities interaction \cite{Zhang2021} was investigated. The new setup can recover in principle all the wave-interaction scenarios that are resolvable on a given discretization,  {which means that robustness may be evaluated thoroughly in all the possible scenarios}.
A simple scalar linear advection equation is taken as the governing equation, by which we are able to clearly measure  overshooting shock-capturing oscillations of high-order numerical schemes, while excluding other factors, e.g., choices over component-wise and characteristic-wise reconstructions \cite{Harten1987}. Moreover, it is necessary for shock-capturing schemes to be reliable first in relatively simpler cases, including the linear equation discussed here, before facing more complicated challenges.    {Of course, more complicated equations can be discussed by having similar initial conditions imposed, but this work intends to provide a simple example for the primary idea, and thus more complicated cases are omitted here for simplicity.}

The remainder of this paper is organized as follows. In the next section, our numerical procedure and its rationale are introduced. In section \ref{sec:results}, numerical test cases involving several well-known shock-capturing schemes are provided to explain the usage of the present approach, and some numerical behaviors are revealed. Finally, concluding remarks are given in the last section.

 \section{The numerical setup} \label{eq:method}

 {To quantitatively evaluate the overshooting oscillations of numerical solutions, we first  define the "overshooting oscillation" or "overshooting error". For any given one-dimensional numerical solution $\hat{u}(x, t)$ of the corresponding exact  solution $u(x, t)$ at a given time $t$, the numerical error at each spatial grid point $x_i$ can be calculated as
\begin{equation} 
u_{\epsilon}(x_i, t)=\hat{u}(x_i, t)-{u}(x_i, t).
\end{equation} 
\noindent However, using this formula indicates that both amplification and attenuation are considered, but here we only discuss  amplification of  numerical solutions, as attenuation typically does not cause but reduces numerical oscillations. Therefore, we redefine the overshooting oscillatory error as 
\begin{equation}\label{overshoot}
u_{\epsilon}(t)= \max _{ 0 \le i\le N }\Big( |\hat{u}(x_i, t)|-|{u}(x_i, t)|, 0 \Big),
\end{equation}
\noindent where $N+1$ is the number of grid points used in numerical tests. Using this formula indicates that diffusive effects that reduce the amplitudes of  waves are not considered in the following discussion. Moreover, here only the maximum error on the given discretization is discussed, as smaller numerical oscillations may be easily smoothed out by numerical dissipation in realistic simulations and thus do not necessarily damage robustness.}

As briefly mentioned in the Introduction, we consider an initial-boundary value problem (IBVP) of one-dimensional wave propagation in an
unbounded domain, governed by the one-dimensional scalar linear advection equation
\begin{equation} \label{eq:hcl}
\frac{{\partial u}}{{\partial t}} + \frac{{\partial u}}{{\partial x}} =0, \quad -\infty <x <+ \infty.
\end{equation} 
\noindent Here, as the equation is linear and has a constant advection speed 1, we have removed potential nonlinear effects that could be caused by nonlinear governing equations. In other words, any nonlinear behavior in the numerical tests below appears only because of nonlinear numerical schemes imposed on nonlinear initial solutions. Certainly, nonlinear equations can be discussed as well, but we can already find interesting information based on the current choice.

To mimic idealized discontinuity/smooth-wave interactions based on the governing equation, each numerical test is initialized by a monochromatic square wave and a  monochromatic smooth sine wave superposing each other, while each wave is characterized by a given  wavenumber. The initial solutions are described by
\begin{equation} \label{eq:linear}
 u(x,0)= \text{sign}\left[\text{cos}(\omega_{\text{sq}} \, x)\right] + A \text{cos}(\omega_{\text{sm}}\, x),
 \label{eqn:soln2}
\end{equation}
\noindent where two wavenumbers  are related to wavelengths by $\omega_{\text{sq}}=2\pi/\lambda_{\text{sq}}$, and $\omega_{\text{sm}}=2\pi/\lambda_{\text{sm}}$, respectively. Then the first term (with subscript "sq") on the right-hand side of Eq.(\ref{eq:linear}) describes \textbf{sq}uare waves, which mimic discontinuities. The coefficient  $A$ is introduced to model the relative difference between the amplitudes of  \textbf{sm}ooth waves (described by the second term on the right-hand side of Eq.(\ref{eq:linear}), with subscript "sm") and discontinuities. This parameter is introduced because weak discontinuities, which may have similar amplitudes as  the smooth waves,  are difficult to numerically distinguish and may still cause numerical oscillations. Here, the exact solutions of this linear IBVP are given by $u(x,t)= u(x-t, 0)$, as the constant characteristic speed is 1.

The spatial discretization of Eq.\eqref{eq:hcl} is performed on a uniform grid, while the coordinate of the $i-$th grid point of total $N+1$ grid points is $x_i = i \Delta x$. Periodic boundary conditions are given at the left ($x=0$) and right ($x=L$) boundaries to model the unbounded domain. Numerical tests below have  $L=1$, while $N$ is given as 500 to provide (more than) sufficient information. 
The supported modes of  square waves and sine waves on the given grid are represented by 
\begin{equation}
\lambda_{\text{sq},m} = L/m, \quad  \text{and } \quad  \lambda_{\text{sm},n} = L/n,   
\end{equation}
respectively, where $m,n = 1,\cdots, N/2$. The corresponding reduced wavenumbers are defined as 
$\varphi_{\text{sq},m} = (2\pi/\lambda_{\text{sq},m})\Delta x $, and $\varphi_{\text{sm},n} = (2\pi/\lambda_{\text{sm},n})\Delta x $, respectively. Then the reduced wavenumbers  lie continuously in $[0,\pi]$ for the rest of the analysis. By including all the supported wave modes, we are able to recover all the possible one-dimensional smooth wave-discontinuity interaction scenarios, providing a complete examination of nonlinear shock-capturing schemes, which may produce numerical oscillations in the vicinity of discontinuities. In particular, overshooting oscillations may be found for certain wavenumbers, thus providing information for understanding the robustness of  corresponding schemes. Whereas, in conventional numerical tests \cite{Lax1998}, wave structures are typically captured on a few selected meshes, which means that only a limited set of reduced wavenumbers is included while examining important flow features. Consequently, the robustness of one or a few simulations does not guarantee the same level of robustness in other test cases. 

We use the third-order strongly stable Runge-Kutta method \cite{Gottlieb2001} for temporal solutions. Changing the temporal solution scheme may also affect the numerical oscillations, but we focus on the relative differences between different spatial approximations done by shock-capturing schemes mentioned in the next section. If different temporal solution schemes are of interest, they can also be compared based on the present approach. Furthermore, it is shown that numerical diffusivity in long-time simulations may inevitably smooth out all the discontinuities and smooth waves,  even when using high-resolution schemes \cite{Deng2019}. Also, in long-time simulations, numerical oscillations may propagate and thus it is difficult to locate the origins of the oscillations. Therefore,  to minimize the inherent diffusive effect of shock-capturing schemes, numerical computations are performed for only one fixed time step  $\Delta t$,  which  is given based on $\text{CFL} = 0.6$. Changing the time step may cause different overshooting oscillations \cite{Zhang2021}, but here it is not further discussed for simplicity. However, it is worth noting that while around CFL=0.6, both the fifth-order WENO-JS scheme and fifth-order WENO-Z scheme show relatively low overshooting oscillations \cite{Zhang2021}, even compared with some results produced with smaller CFL numbers. Although for the seventh-order WENO schemes, the behavior is not exactly the same \cite{Zhang2021}, the same CFL number is still used for a more consistent comparison. Of course, the CFL number can be easily changed, but the essential idea of the present study, which is about the quantification of overshooting oscillations, is still valid.
Finally, the behaviors of $u_{\epsilon}(t)$ with respect to $\omega_{\text{sq}}$, $\omega_{\text{sm}}$, and $A$ are discussed below.

 \section{Numerical examples and discussions} \label{sec:results}%
In this work, the previous IBVP is solved by a series of selected representative shock-capturing schemes, of which the details are not discussed here. 
The  schemes selected here include the total variation diminishing (TVD) schemes \cite{Harten1984,Sweby1984}, which are known for their robustness,  essentially non-oscillatory (ENO) schemes \cite{Harten1987,Shu1989}, (monotonicity preserving) weighted essentially non-oscillatory ((MP)WENO) schemes \cite{Liu1994,Jiang1996,Balsara2000,Borges2008}, which are commonly used in numerous applications, and WENO-CU6 scheme \cite{HUXY_2010}, which is a central-upwind scheme that includes a six-point candidate stencil, in addition to the typical three-point sub-stencils. 
 With the given numerical setup,  TVD schemes and the third-order ENO scheme (ENO3) do not produce overshooting oscillations, and the oscillations produced by the fifth and seventh-order MPWENO schemes are significantly smaller than those of the fifth-order WENO-JS scheme \cite{Jiang1996}.  Therefore, corresponding (almost) oscillation-free results are not further discussed. Note that when the CFL number becomes larger ($>$0.8), some of these relatively robust schemes (e.g., the TVD-superbee scheme \cite{Sweby1984}) might produce weak overshooting oscillations \cite{Zhang2021}, but their overall robustness is still obvious.
  Moreover, although it is already known that some shock-capturing schemes are more robust for challenging test cases \cite{Balsara2000,Zhao2019}, below the differences between different shock-capturing schemes and the scenarios that cause overshooting oscillations are quantitatively and also more comprehensively shown, thus with more insights provided.

\begin{figure}[htbp]
\centering 
\includegraphics[width=0.49\textwidth]{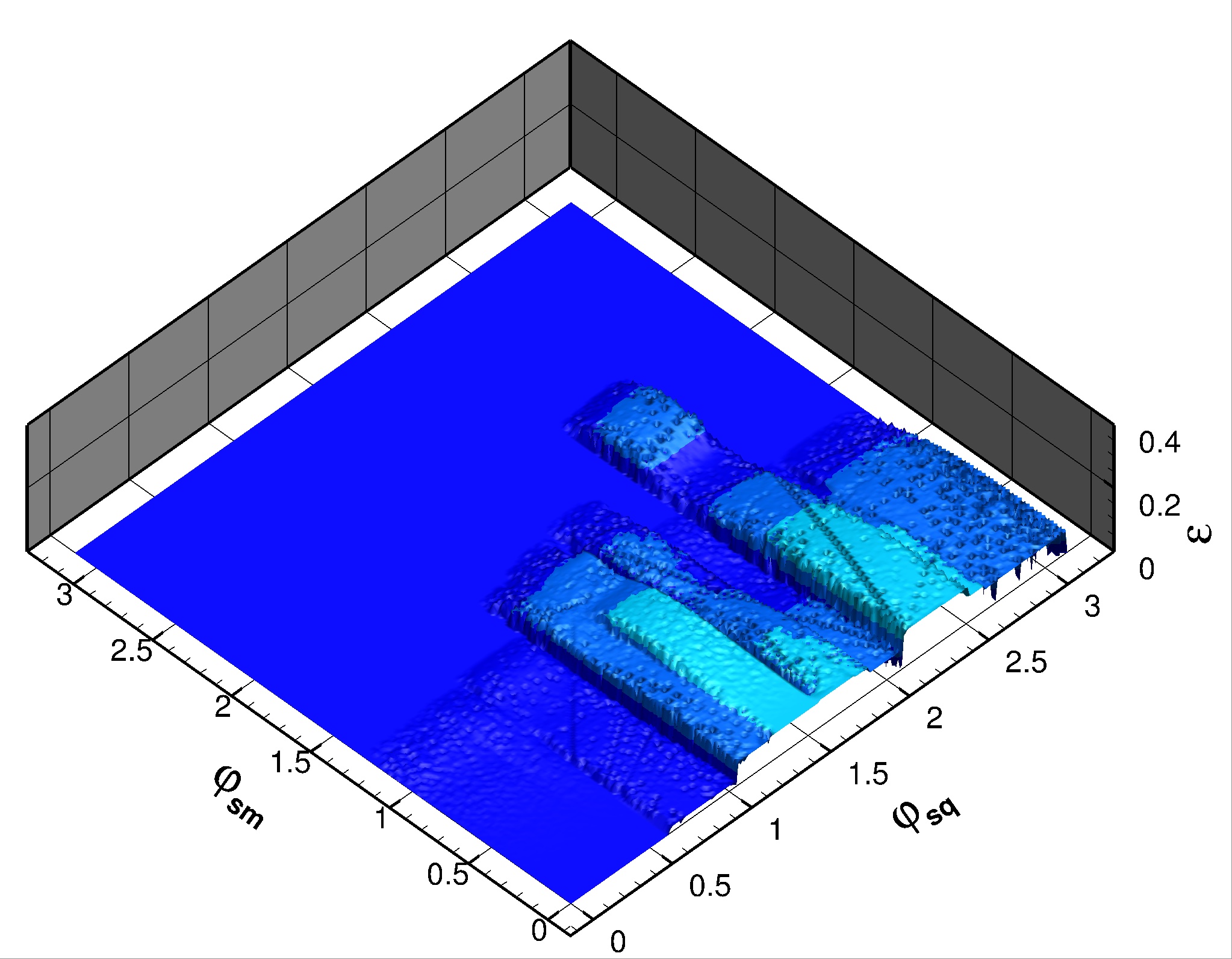}\\{(a) A=0.4}
\includegraphics[width=0.49\textwidth]{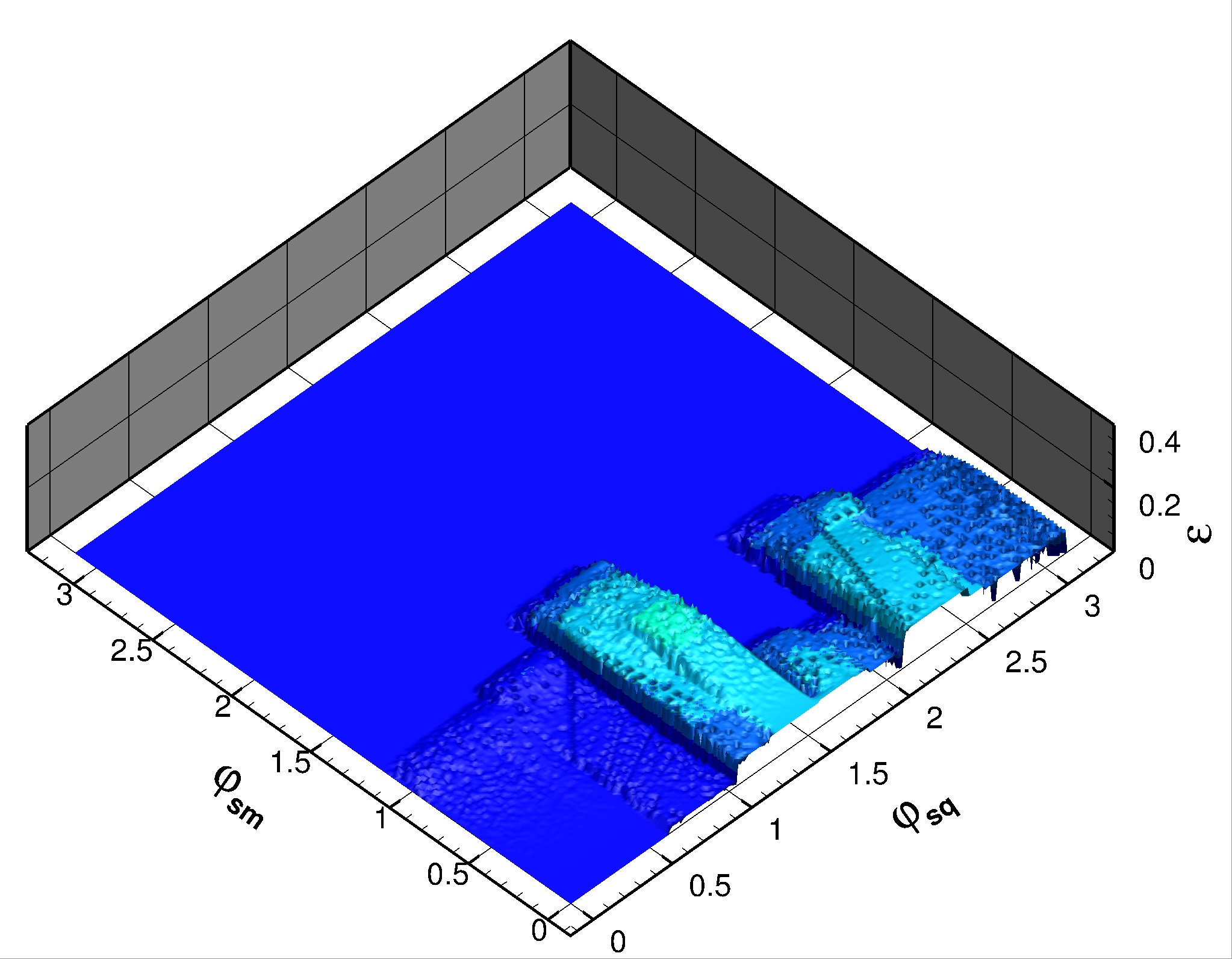}\\{(b) A=0.8} 
 \caption{Overshooting errors of the WENO-JS5 scheme.}
 \label{fig:db5}
\end{figure}

 Firstly, the results of the commonly used fifth-order WENO-JS scheme (WENO-JS5) are shown in Fig. \ref{fig:db5}. The WENO-JS5 scheme produces much smaller overshoots than other fifth-order (W)ENO-type schemes (except MPWENO schemes that have an extra limiting procedure), which explains its robustness in complex compressible flow modeling \cite{Larsson_2013,Zuo_2019}. However, relatively minor overshoots are still produced if the discontinuities are close. In particular, when $A$ is larger, the overshooting oscillations  increase when the wavenumbers lie in a certain range (which also means that the overshooting oscillations can be stronger without needing to have stronger discontinuities). For instance, if $\varphi_\text{sq} \in (1.2, 1.5)$ and $\varphi_{\text{sm}}\in (0.7, 1.0)$ (approximately), the overshooting oscillations are stronger. Note that within the range of $\varphi_\text{sq} \in (1.2, 1.5)$, smooth waves can usually be well resolved by upwind finite-difference schemes. However, if discontinuities exist and are closely located, even the robust WENO-JS5 scheme may produce overshooting oscillations. Of course, this behavior is, to some extent, inevitable for high-order schemes as the sub-stencils for spatial reconstructions may cross at least one discontinuity when discontinuities are closely located, but note that TVD schemes, ENO3 scheme, and MPWENO schemes are free from this issue (but more diffusive). Therefore, this behavior suggests that
 attention to closely located multiple discontinuities is needed, and 
the robustness of high-order schemes cannot be evaluated simply based on a few numerical cases that may not fully reproduce the oscillation-prone scenarios.

\begin{figure}[htbp]
 \centering  
 \includegraphics[width=0.49\textwidth]{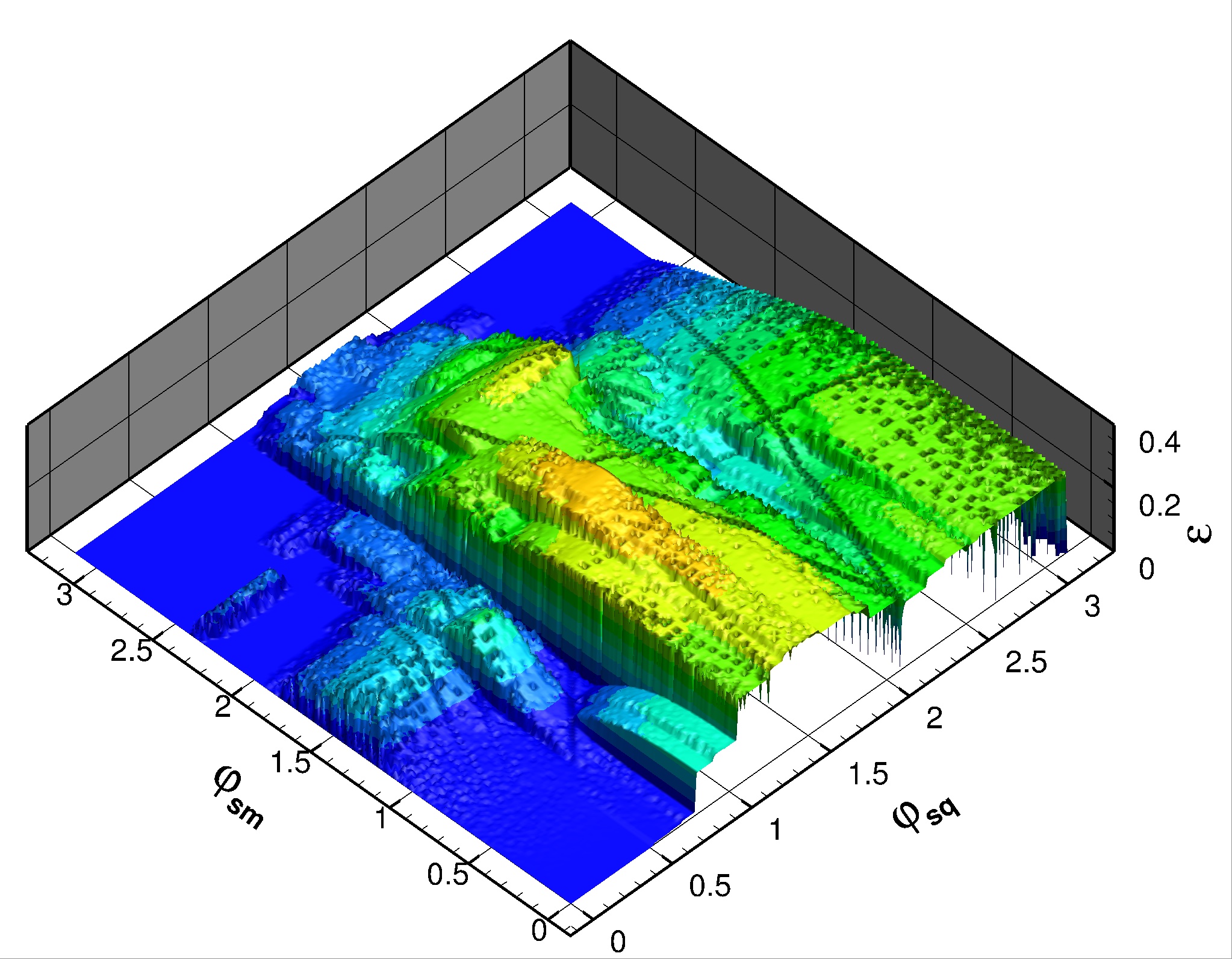}\\{(a) A=0.4} 
 \includegraphics[width=0.49\textwidth]{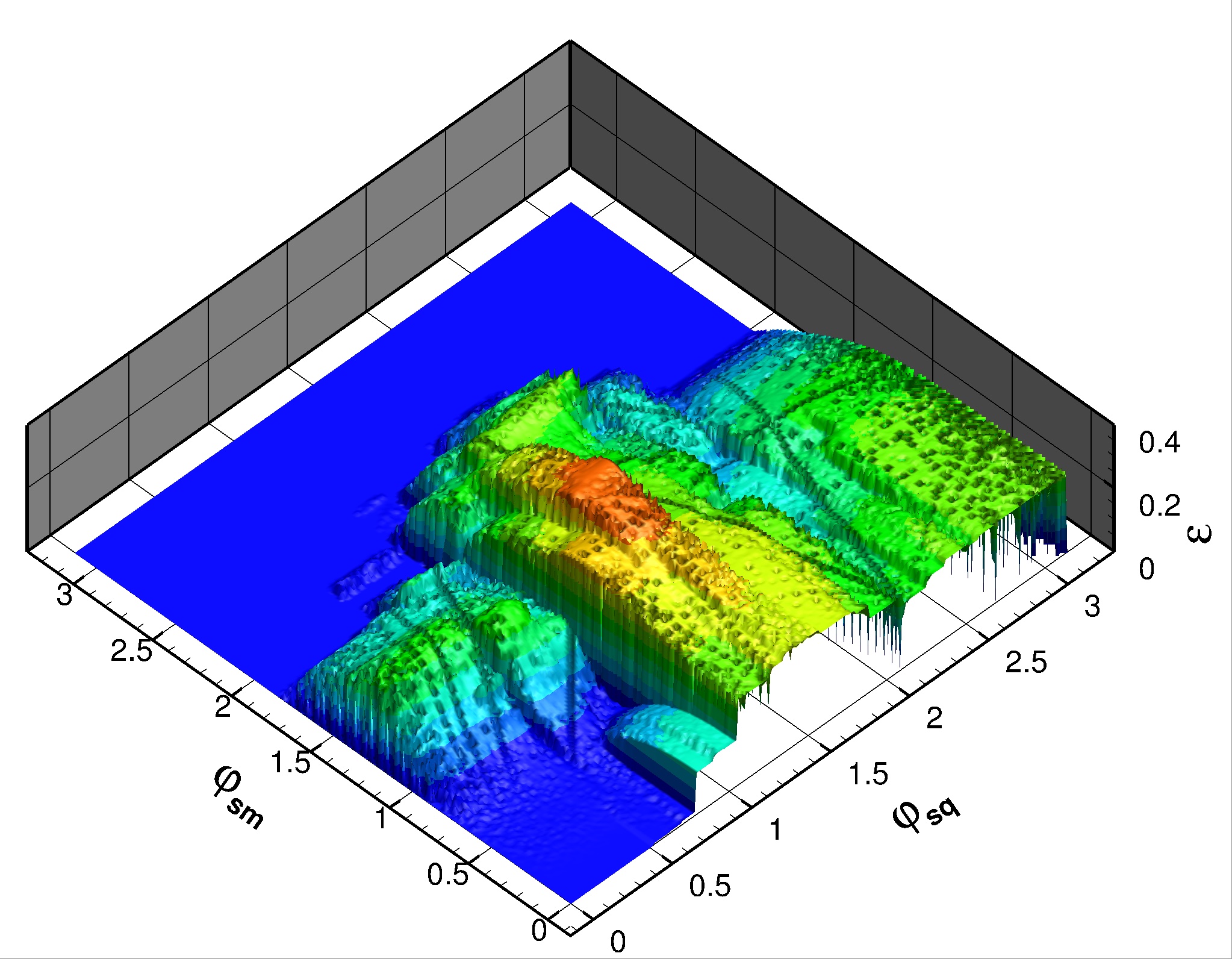}\\{(b) A=0.8} 
 \caption{Overshooting errors of the WENO-JS7 scheme.}
 \label{fig:db6}
\end{figure}

Then, the WENO-JS7 scheme, which uses wider (four-point) sub-stencils, produces more significant numerical overshoots, as shown in Fig.\ref{fig:db6}. Moreover, with $\varphi_\text{sq} \in (0.8, 1.0)$, the WENO-JS5 scheme does not cause significant overshooting oscillations, but the WENO-JS7 scheme already produces overshooting oscillations that are even larger than the maximum overshoot of the WENO-JS5 scheme. An important reason causing this phenomenon is that the four-point sub-stencil may at least cross one discontinuity with $\varphi_\text{sq} \in (0.8, 1.0)$, while the three-point sub-stencils used by the WENO-JS5 scheme could tolerate more closely located discontinuities without crossing them.  {This behavior suggests that, for instance, the incremental-stencil reconstruction \cite{Wang2018} may be useful for WENO-type schemes, as in that case, at least one smaller but smooth sub-stencil is always available even if discontinuities are closely located. However, a detailed discussion is beyond the scope of this work, as the corresponding nonlinear weighting process also needs to be carefully investigated}. Another interesting behavior of the WENO-JS7 scheme is that, while $\varphi_\text{sm} \in (1.0, 1.7)$, crossing a single discontinuity (when $\varphi_\text{sq}$ is small)  still leads to significant overshooting oscillations, especially when $A$ is large. We believe this behavior appears because WENO schemes do not completely discard the oscillatory stencils. In this scenario, the numerically evaluated smoothness of each sub-stencil may be similar, and thus the oscillatory sub-stencil is not sufficiently excluded from the final reconstruction. 

\begin{figure}[htbp]
 \centering 
\includegraphics[width=0.49\textwidth]{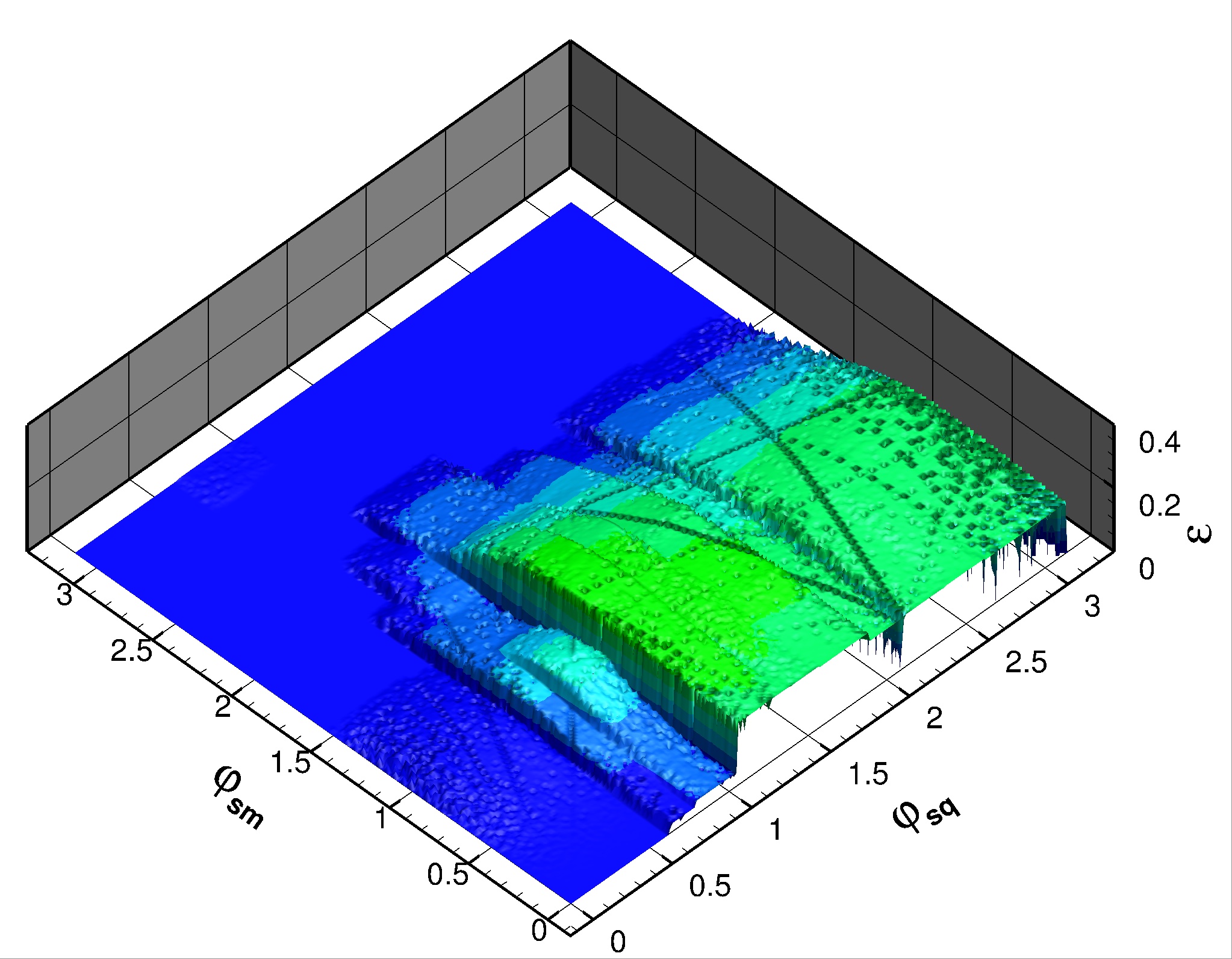}\\{(a) A=0.4} 
\includegraphics[width=0.49\textwidth]{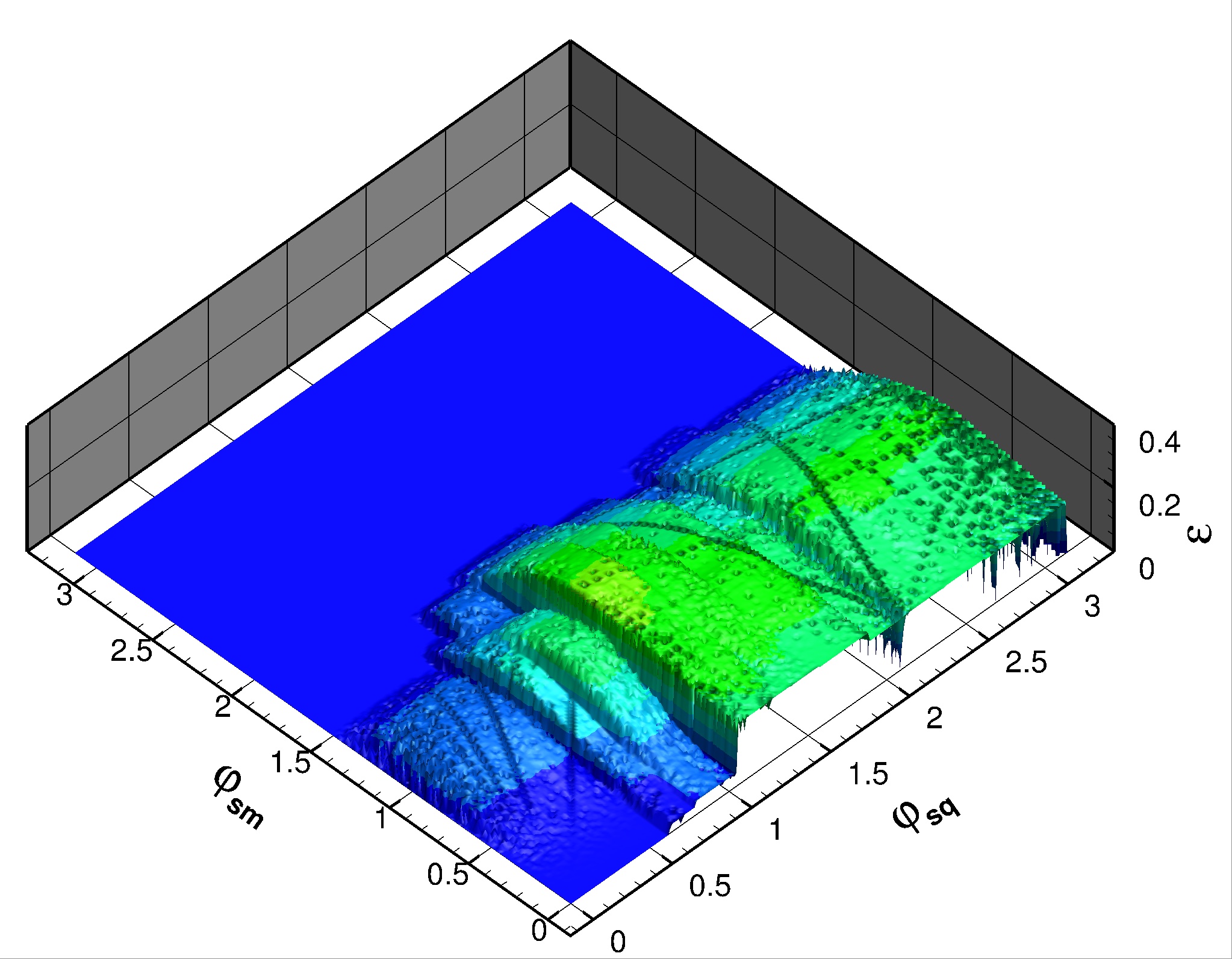}\\{(b) A=0.8} 
 \caption{Overshooting errors of the WENO-Z5 scheme.}
 \label{fig:db7}
\end{figure}

Then the results of the WENO-Z5 scheme are shown in Fig.\ref{fig:db7}. The numerical overshooting oscillations are clearly more significant than those of the WENO-JS5 scheme, which is also fifth-order accurate and has the same sub-stencils. This is not surprising, because, for the WENO-Z5 scheme, non-smooth sub-stencils contribute more to the spatial reconstruction.
It is well-known and easy to verify that the WENO-Z scheme is able to keep the designed accuracy even at (high-order) smooth critical points \cite{Don2013}, but the difference between the robustness of the WENO-JS scheme and the WENO-Z scheme was difficult to quantitatively evaluate, via the conventional approach of using a series of classic test cases \cite{Lax1998}. Here, we can see that for shock-capturing simulations the WENO-JS5 scheme is more robust because its maximum numerical overshooting oscillation is around half of that of the WENO-Z5 scheme. Moreover, the WENO-Z5 scheme also produces overshooting oscillations in wider ranges of wavenumbers. Of course, as mentioned above, it is challenging to retain robustness while accuracy and/or resolution are (significantly) improved, but with a quantitative measurement like the present one, we may be able to evaluate the performance more thoroughly and improve the performance more adequately.  {Based on the present result, it may be important to further improve the nonlinear weighting or impose an extra adaptation procedure, to be able to more effectively detect discontinuities and use more adequate reconstruction schemes respectively in smooth region and near discontinuities.}

\begin{figure}[htbp]
 \centering  
 \includegraphics[width=0.49\textwidth]{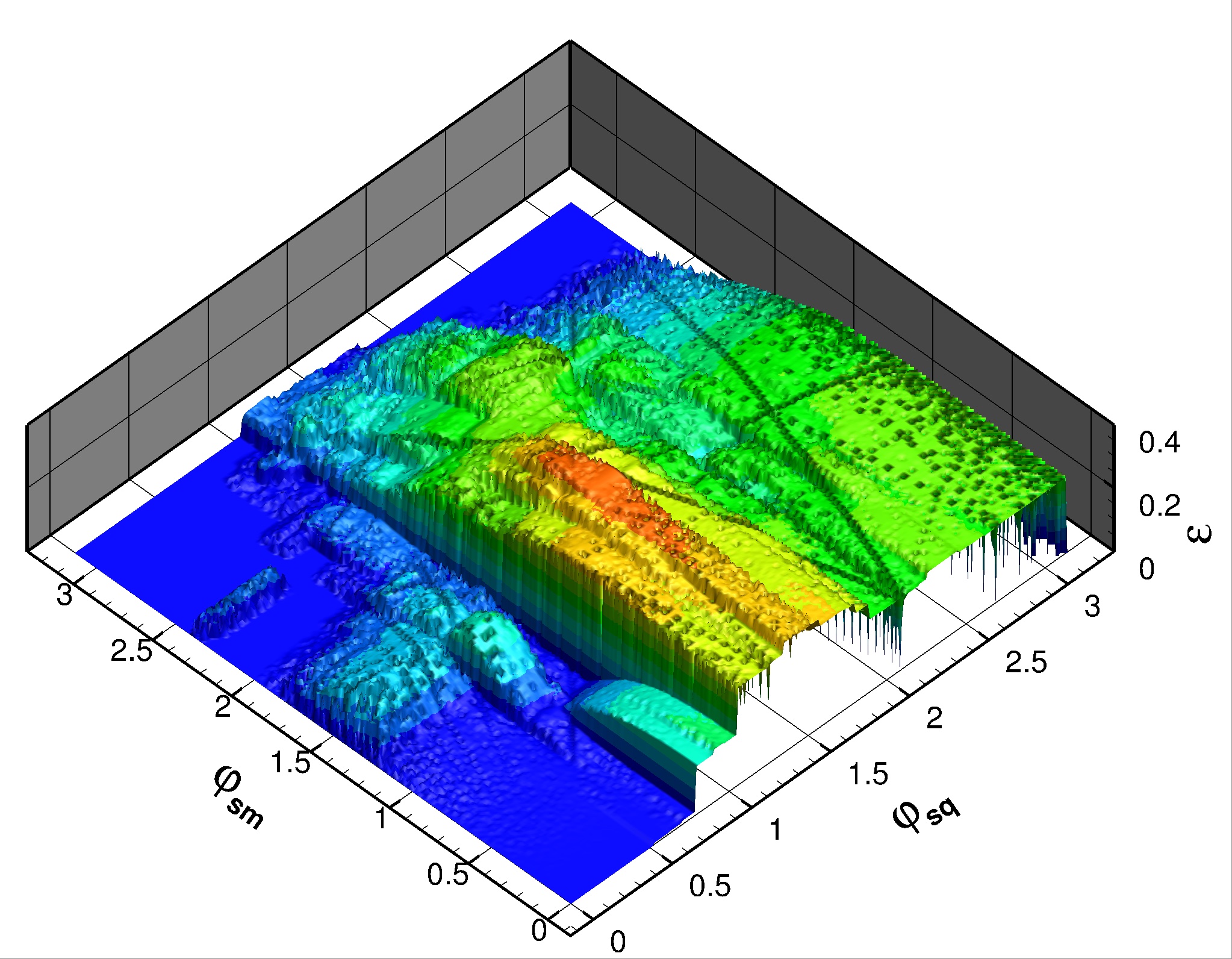}\\{(a) A=0.4} 
 \includegraphics[width=0.49\textwidth]{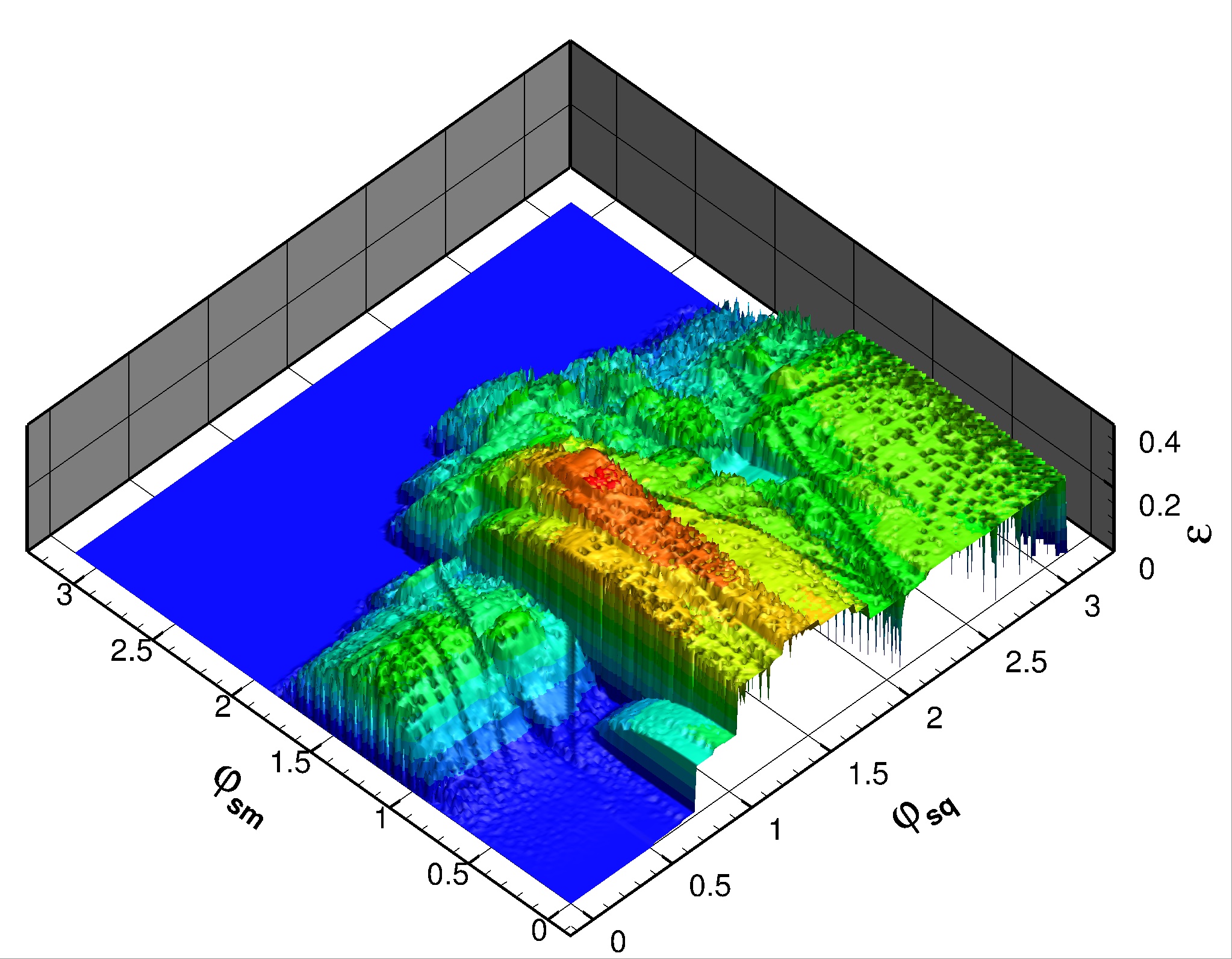}\\{(b) A=0.8} 
 \caption{Overshooting errors of the WENO-Z7 scheme.}
 \label{fig:db9}
\end{figure}

Furthermore, as shown in Fig.\ref{fig:db9}, the numerical overshooting oscillations of the WENO-Z7 scheme are stronger than those of the WENO-Z5 scheme.
In comparison with the WENO-JS7 scheme, the WENO-Z7 scheme also produces stronger oscillations. However, the difference between these two seventh-order schemes is less significant than the difference between two fifth-order schemes. Moreover, the oscillations show a significant nonlinear correlation with $A$ in the wavenumber ranges of $\varphi_\text{sm} \approx 1.5$ and  $\varphi_\text{sq} < 0.7$ (figures for more $A$ values are not shown here). This behavior again suggests that the overshooting oscillations cannot be overlooked even if the shock waves are relatively weak (compared to the smooth waves).  {It is also worth noting that, in the same range of wavenumbers, where the discontinuities are sufficiently separated (meaning that at least one of the sub-stencils is not crossing any discontinuity), the seventh-order schemes are still more prone to cause overshooting oscillations than the fifth-order schemes. Therefore, this issue may need to be addressed while improving the seventh-order schemes.}

The last scheme being tested is the WENO-CU6 scheme 
\cite{HUXY_2010}, which is sixth-order accurate for the linear advection problem \cite{Yamaleev2009}. The overall behavior of the overshooting oscillations of the WENO-CU6 scheme is similar to those of the WENO-JS and WENO-Z schemes, although the WENO-CU6 scheme uses a different way to construct its (sub-)stencil(s). However, when $A$ is relatively small and discontinuities are sufficiently far away from each other (i.e., having small $\varphi_{\text{sq}}$), the WENO-CU6 scheme may produce relatively weak but noticeable overshooting oscillations when the wavenumbers of smooth waves are high. 
This behavior is not observed for other WENO-type schemes, and it means that extra attention might be needed if the WENO-CU6 scheme is used for capturing relatively weak but high-frequency smooth waves. 
More importantly, without having a complete set of wave modes included, this behavior was also difficult (if not impossible) to observe via conventional numerical tests. 

\begin{figure}[htbp]
\centering 
\includegraphics[width=0.49\textwidth]{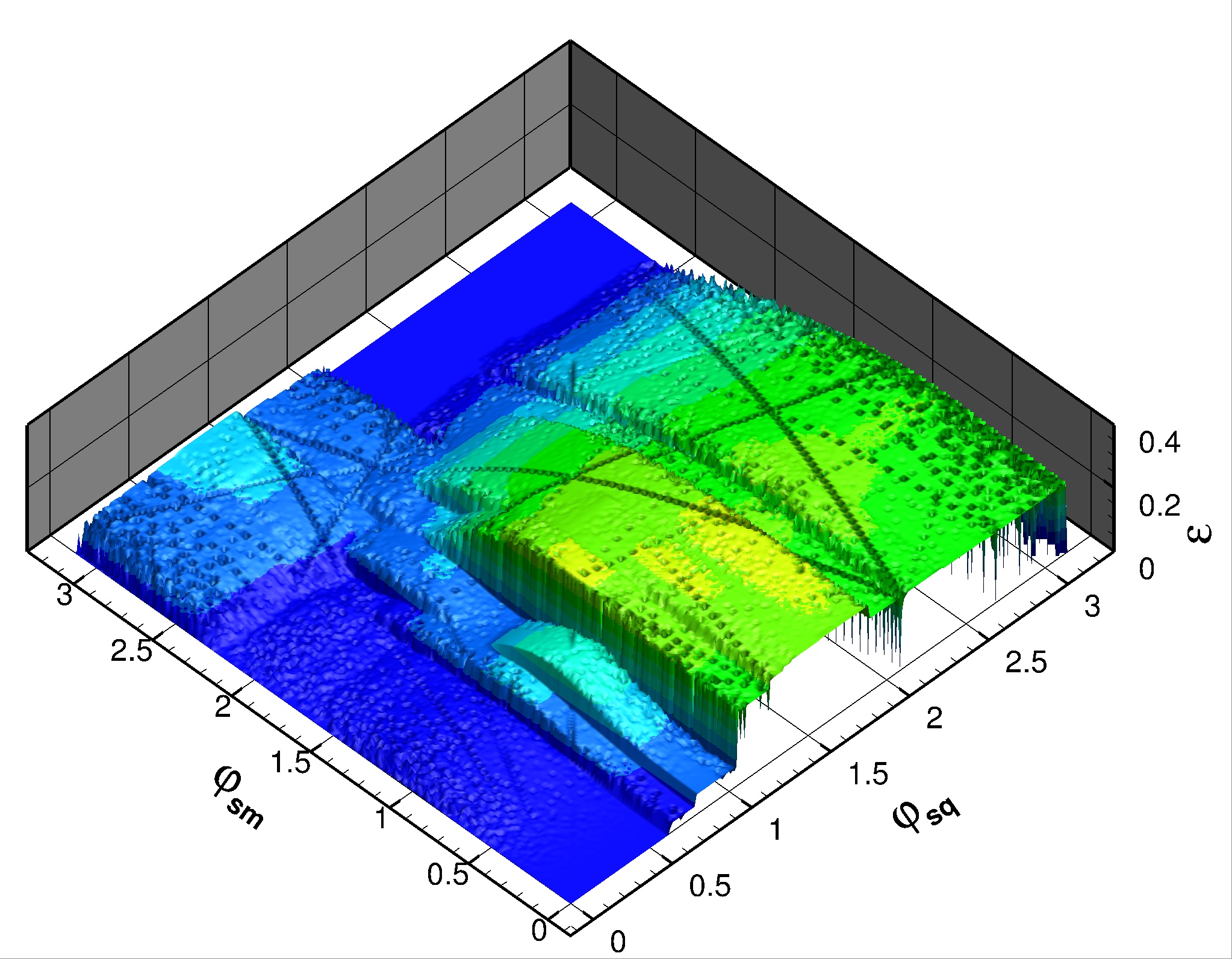}\\{(a) A=0.4}
\includegraphics[width=0.49\textwidth]{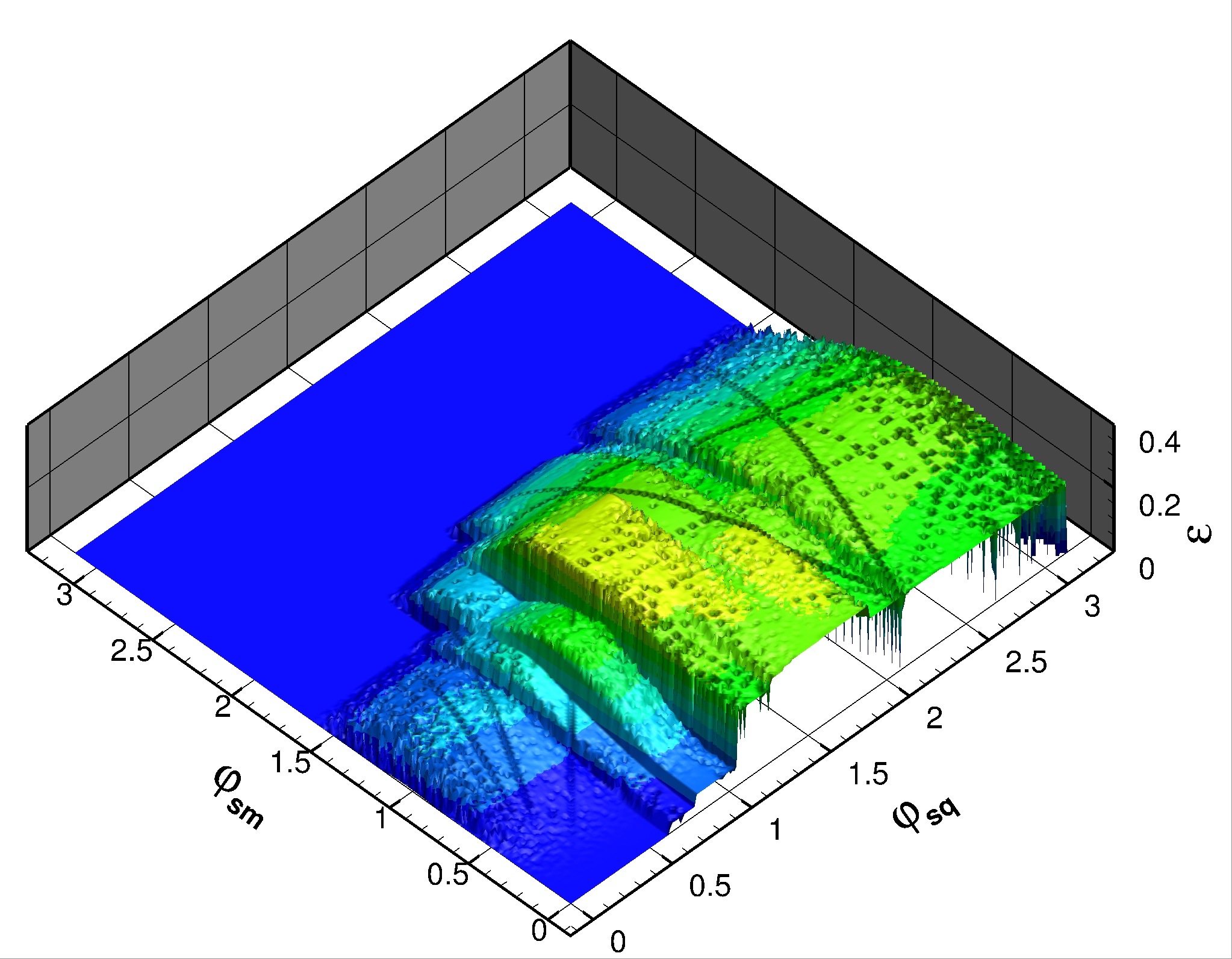}\\{(b) A=0.8} 
 \caption{Overshooting errors of the WENO-CU6 scheme.}
 \label{fig:db11}
\end{figure}

\begin{figure}[htbp]
 \centering  
 \includegraphics[width=0.49\textwidth]{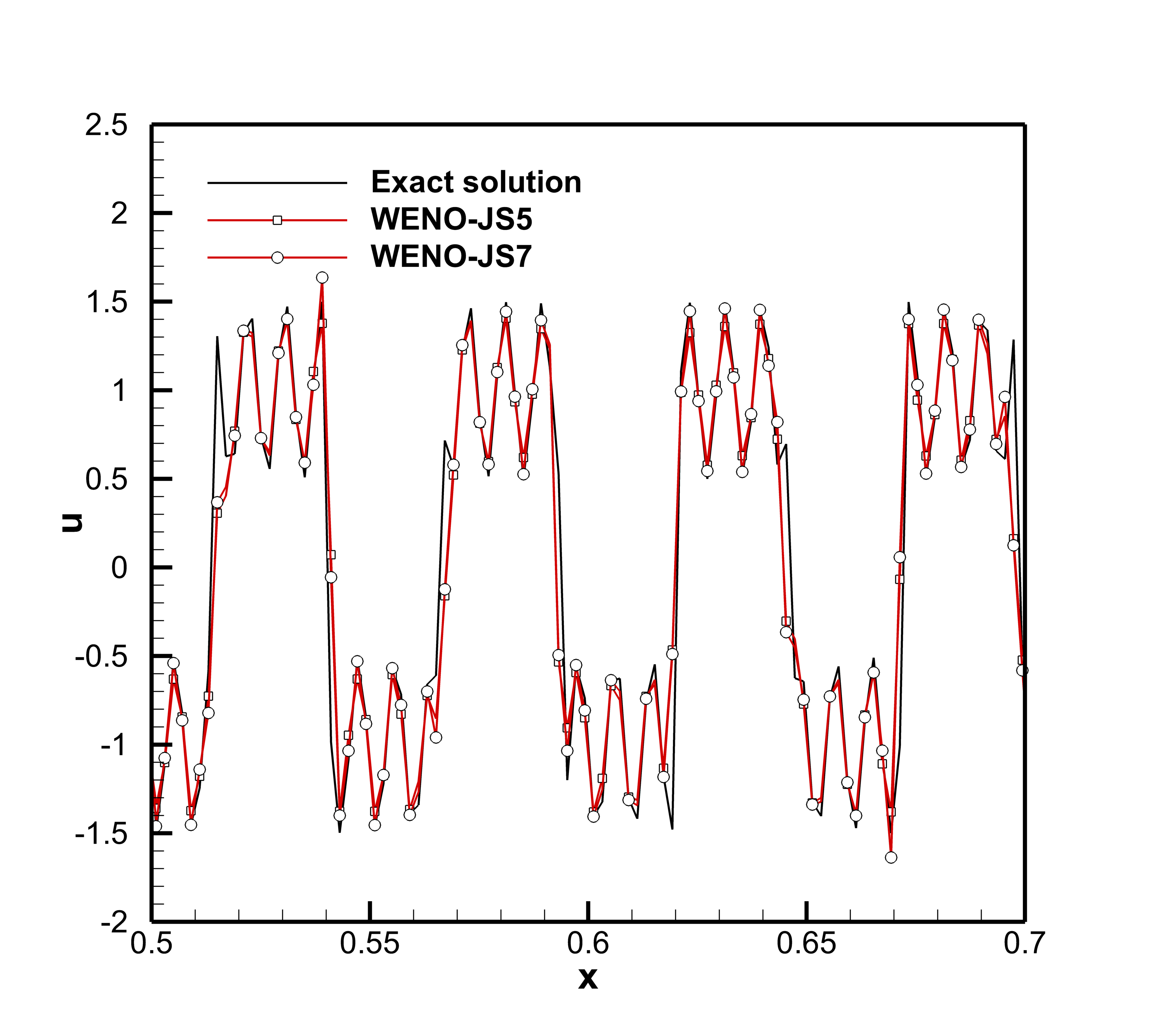}\\{(a) WENO-JS} 
 \includegraphics[width=0.49\textwidth]{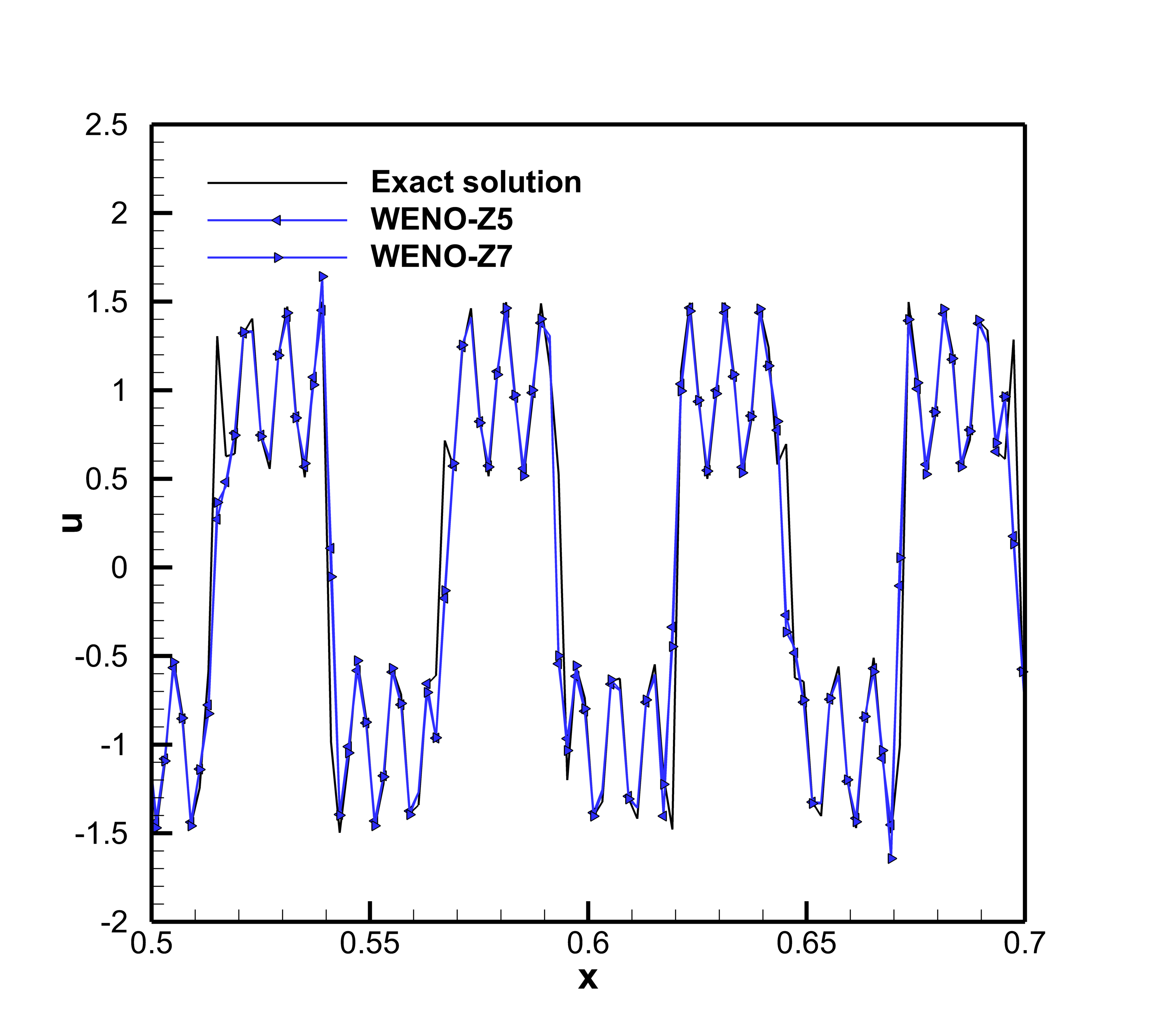}\\{(b) WENO-Z} 
 \caption{Local solution distributions for $\varphi_{\text{sq}}$=0.24 and $\varphi_{\text{sm}}$=1.50.}
 \label{fig:db10}
\end{figure}

Finally, to provide a different perspective for the present results, some spatial solution distributions are shown in Fig. \ref{fig:db10}. In these results, the distance between discontinuities is large enough to ensure that at least one of the sub-stencils of WENO-type schemes does not cross any discontinuity.  The significant overshooting oscillations of the WENO-JS7 and WENO-Z7 schemes are found in the  vicinity of discontinuities. Note that if the wavenumber of the smooth wave further reduces to $\varphi_{\text{sm}}<1$, the overshooting oscillations of the fifth-order schemes may increase, but the overshooting oscillations of the seventh-order schemes may decrease, as shown in previous figures. These behaviors suggest that for an isolated shock interacting with (relatively) lower frequency smooth waves, a higher-order scheme (WENO-Z7 vs. WENO-Z5) could be (slightly) more robust than a lower-order scheme, which is contradictory to the general assumption. Of course, a (relatively) lower-order accurate scheme is still more reliable in most of the scenarios, but this example again shows that the present measurement can more systematically and quantitatively evaluate the robustness of different shock-capturing schemes.

  \section{Conclusions} 
We have designed a simple approach to evaluate overshooting shock-capturing oscillations produced by nonlinear numerical schemes.
Based on the obtained numerical tests, several direct observations can be given: (I) the amplitudes of overshooting oscillations
significantly depend on the reduced wavenumbers of smooth waves and square waves, while high-wavenumber square waves represent closely located discontinuities;  (II) high-resolution-low-dissipation schemes  produce stronger oscillations, in comparison with their lower-order-accurate counterparts; and (III) the MPWENO5 scheme is the only high-order scheme that provides (almost) overshoot-free results in most of the scenarios. As these observations are mostly expected according to general numerical experience, we provide quantitative information, which was not fully available before.

More importantly, to evaluate the robustness of shock-capturing schemes,  we no longer (only) rely on the typical qualitative comparison of numerical tests in a case-by-case manner. Note that in a complex and more realistic simulation,
it is difficult to observe the origins and magnitudes of numerical oscillations. Whereas, the present procedure can systematically and quantitatively show the overshooting oscillations and their dependencies on  wavenumber, wave amplitude, and the distance between discontinuities.  In particular,
the present approach involves a complete set of wave modes supported by the given discretization, thus recovering complete wave-interaction scenarios that may exist in realistic simulations, while using a limited number of numerical test cases (shock tube problems, 2D Riemann problems, etc.) may overlook certain scenarios. In fact,  as shown in the last section,  having the complete sets of wave modes allows us to observe some delicate variations that were difficult to observe in conventional numerical tests. 

 {Like our previous work \cite{Zhang2021}, the present approach can also be used for evaluating available and newly developing shock-capturing schemes, as shown in Ref. \cite{Tang2023}, providing a new perspective to understand the performance (particularly, the robustness) of sophisticated nonlinear schemes. The finding can also be useful for designing new shock-capturing schemes. For instance, when the discontinuities are getting close, having smaller sub-stencil(s) in the shock-capturing schemes may be beneficial to delay the appearing of overshooting oscillations, as the small sub-stencil(s) may be able to avoid crossing  any discontinuity. Of course, as already discussed above, designing a new shock-capturing scheme needs to consider both robustness and  accuracy. The present approach provides a new tool  for evaluating the robustness (only), which was a weakness in achieving a comprehensive analysis of numerical schemes.}

\section*{Data Availability Statement}
The data that support the findings of
this study is available from the
 author upon reasonable
request.
\section*{Acknowledgments} 
The author thanks Huaibao Zhang, Chunguang Xu, and Guangxue Wang, for valuable discussion and support.

 
\bibliographystyle{unsrt} 
\bibliography{sample1}
\end{document}